# Unconventional and Fragile Magnetic Exciton in a van der Waals Quantum Magnet

Kai-Xuan Zhang[1†*], Min Zhang[2†], Minjae Kim[3,4†], Yong-Hyun Kim[5], Junghyun Kim[1], Heejun Yang[1], Pyeongjae Park[1], Chaebin Kim[1], Mangesh Diware[1], Junik Hwang[6], Youjin Lee[1], Byeong-Gwan Cho[7], Hyeong-Do Kim[7], Tae-Yeong Koo[7], Chunhua Chen[8], Mingtao Li[8], Xujie Lü[8], Wenge Yang[8], Kee-Hoon Kim[1], Seung-Ho Baek[6], Hyeonsik Cheong[9,10], Sung-Keun Lee[5], Beom Hyun Kim[1], Christopher Lane[11,12], Jian-Xin Zhu[11,12], Zhaorong Yang[2,13*], Young-Woo Son[3*], Je-Geun Park[1*]

[1]*Department of Physics and Astronomy, Seoul National University, Seoul, 08826, South Korea*

[2]*Institutes of Physical Science and Information Technology, Anhui University, Hefei, 230601, China*

[3]*Korea Institute for Advanced Study, Seoul, 02455, South Korea*

[4]*Department of Semiconductor Science & Technology, Jeonbuk National University, Jeonju, 54896, South Korea*

[5]*School of Earth and Environmental Sciences, Seoul National University, Seoul, 08826, South Korea*

[6]*Department of Semiconductor Physics, Changwon National University, Changwon, 51139, South Korea*

[7]*Pohang Accelerator Laboratory, Pohang, 37673, South Korea*

[8]*Center for High Pressure Science and Technology Advanced Research, Shanghai, 201203, China*

[9]*Department of Physics, Sogang University, Seoul, 04107, South Korea*

[10]*Center for Nano Materials, Sogang University, Seoul, 04107, South Korea*

[11]*Theoretical Division, Los Alamos National Laboratory, New Mexico, 87545, USA*

[12]*Center for Integrated Nanotechnologies, Los Alamos National Laboratory, New Mexico, 87545, USA*

[13]*Anhui Key Laboratory of Low-Energy Quantum Materials and Devices, High Magnetic Field Laboratory, HFIPS, Chinese Academy of Sciences, Hefei, 230031, China*

† These authors contributed equally to this work

* Corresponding authors: Kai-Xuan Zhang (email: kxzhang.research@gmail.com), Zhaorong Yang (email: zryang@issp.ac.cn), Young-Woo Son (email: hand@kias.re.kr), and Je-Geun Park (email: jgpark10@snu.ac.kr)

**Excitons and their complexes, generated by optical excitations in materials, represent fundamental quantum phenomena essential to energy transport and information processing in both natural and engineered systems[1,2]. The recently discovered magnetic exciton in the van der Waals (vdW) antiferromagnet $NiPS_3$ exemplifies these phenomena, exhibiting several distinctive characteristics. Despite extensive investigation[3-10], much of its physics remains unresolved[3-16], with key questions about why the $NiPS_3$ magnetic exciton is so sharp and optically bright despite the nominally spin-forbidden transition, posing significant challenges to a proper understanding and practical manipulation of the exciton. An urgent question is to what extent it is due to chemical disorder, magnetic weakening, lattice modification, or intrinsic instability of the bright exciton itself: answers to which will put**

**stringent constraints on possible theoretical models. Here we address these questions using hydrostatic pressure as a clean, continuous, reversible, and in-situ tuning parameter. We find that the sharp photoluminescence peak is drastically suppressed by as little as 0.4 GPa and completely quenched by 1.5 GPa, with demonstrating its reversibility. Crucially, this bright-to-dark conversion occurs without magnetic, crystallographic, or electronic reconstruction despite an increase in the Neel temperature, as established by Raman, X-ray absorption, nuclear magnetic resonance spectroscopy, and first-principles many-body calculations[13]. Our results demonstrate that the optical brightness of the magnetic exciton is independent of chemical disorder, lattice expansion, and weakening of magnetic order, indicating that a higher-order correlated mechanism governs the bright exciton. We further propose experimentally constrained microscopic scenarios involving exciton pairing, crystal-field-controlled spin-orbit mixing, and symmetry breaking, providing a framework for future tests of entangled magnetic exciton in correlated quantum magnets. Our findings indicate that the sharp coherence of the entangled magnetic exciton benefits from its delicate quantum nature, and its optical brightness can be activated by exciton pairing or spin-orbit coupling, while its fragility under external control is achievable through a high-order perturbation rather than a first-order transition.**

Magnetism, a cornerstone of many-body quantum mechanics, has driven numerous scientific breakthroughs over the decades. The advent of magnetic van der Waals (vdW) materials has opened exciting opportunities for exploring magnetism in genuinely two-dimensional systems[17-21]. Among these materials, transition-metal phosphorus trisulfide[17,18,22,23] stands out as an antiferromagnetic (AFM) insulator that hosts diverse model Hamiltonians, ranging from Ising to Heisenberg models.

Beyond their fundamental significance, vdW magnets have proven instrumental in various technological applications[24,25], particularly in spintronics. Examples include giant spin filter effects[26-28], tunnelling magnetoresistance effects[29,30], spin-orbit torque[31-37], and skyrmions[38-40]. Additionally, vdW magnets have emerged as promising platforms for novel light-matter interactions, enabling the discovery of light-induced out-of-equilibrium dynamics such as Floquet engineering[23], ultrafast exciton dynamics[41], and coherent magnon generation[42-45]. These findings establish vdW magnets, particularly those with spin-orbit coupling, as fertile grounds for exploring unconventional light-matter interaction phenomena.

$NiPS_3$ exhibits a zigzag AFM ground state below its Neel temperature ($T_N$ ~150 K) and hosts several intriguing properties[3-9,17]. Its AFM order is an XXZ-type, or to the lowest order, two-dimensional XY Hamiltonian[3,4], with no stable magnetic order in the monolayer limit[4] consistent with the theoretical prediction of the Berezinskii−Kosterlitz−Thouless transition. This material also demonstrates strong coupling between electronic and magnetic states, as evidenced by charge-spin correlations in optical studies[5]. Moreover, $NiPS_3$ manifests coherent, resolution-limited excitons below $T_N$, originating from transitions between Zhang-Rice triplet (ZRT) and Zhang-Rice singlet (ZRS) states[6]. These excitons exhibit inherently many-body and quantum-entangled behaviour[6], with their coupling intrinsically tied to spin-dependent charge hopping[6]. Many-body calculations further reveal that exchange correlations (Hund's coupling) between self-doped holes induce local dipole polarisation (~0.1-0.2 μC/cm$^2$ per $NiS_6$ octahedron), unveiling a new type of multiferroic physics.

The magnetic exciton in $NiPS_3$ is distinct among 2D vdW magnets due to its sharp photoluminescence (PL) peak at 1.4756 eV, observed only below its $T_N$ of 150 K. As the temperature decreases, the linewidth narrows significantly, achieving a coherence below 50 K with a resolution-limited linewidth of less than 0.4 meV[6]- ten times smaller than the thermal energy of 50 K. This unique feature is complemented by other unusual properties, such as spin-induced linear polarisation of PL[7], highly anisotropic excitons, and multiple phonon-bound states[8]. Furthermore, ultrafast terahertz spectroscopy has revealed exciton-driven AFM metallic behaviour[9]. Despite these exciting advances, the fundamental mechanism underlying the PL peak remains unresolved, as the ZRT-to-ZRS transition is optically forbidden due to spin-state variations.

Another defining characteristic of $NiPS_3$ is its strongly correlated Hubbard physics characterised by a large on-site Coulomb interaction of $U$. Optical studies reveal significant spectral weight transfer across a broad energy range (1-6 eV) upon cooling from room temperature to 2 K[5], underscoring its strongly correlated nature. This combination of large Hubbard energy and low dimensionality makes $NiPS_3$ an ideal system for exploring how external perturbations, such as pressure, can modify its physical parameters and ground states. Previous chemical-substitution studies[16,46] revealed that the magnetic exciton in $NiPS_3$ is suppressed by cation or anion doping. They also left an important ambiguity unresolved: Chemical substitution simultaneously introduces local disorder and inevitably compares different chemically modified crystals. Therefore, it cannot determine whether the exciton suppression is caused by extrinsic disorder, weakened magnetic correlations, lattice expansion, or an intrinsic instability of the optical brightening mechanism.

On the other hand, hydrostatic pressure provides a fundamentally different, clean test because it tunes the same crystal continuously and reversibly without chemical substitution. In particular, pressure allows us to check whether the optical brightness of the magnetic exciton can be destroyed even when the lattice is compressed, and the antiferromagnetic transition temperature is enhanced. However, the effect of pressure on its magnetic exciton remains largely unexplored and could provide an effective, clean means of manipulation. Unravelling how pressure mediates these interactions could also provide critical insights into the unresolved puzzles surrounding $NiPS_3$ and unlock a deeper understanding of this fascinating material.

We report that the PL amplitude of the exciton in $NiPS_3$ decreases significantly by 80% with increasing pressure up to 0.4 GPa and vanishes entirely at 1.5 GPa. This dramatic suppression occurs without significant change in the material's structural, electronic, and magnetic properties. The magnetic exciton's hypersensitivity to pressure is in stark contrast to the behaviour of Raman peaks, which show no significant variation up to 1.5 GPa, indicating minimal lattice compression, if any, structural change under pressure. Further insights come from high-pressure X-ray absorption spectroscopy (XAS), which reveals no detectable pressure-induced changes in the electronic structure up to 3.2 GPa. This robustness rules out alterations to the electronic ground states under pressure. Complementing these findings, nuclear magnetic resonance (NMR) measurements show a slight, monotonic increase in the Neel temperature ($T_N$) with increasing pressure, suggesting a strengthening of the magnetic ground states. This observation sharply contrasts with the hypothesis that increasing pressure would weaken the magnetic ground states, thereby suppressing PL. Together, these factors establish $NiPS_3$ as a model system in which pressure unveils emergent physics governing the behaviors of its magnetic excitons under stringent experimental constraints and enables remarkably effective clean manipulation.

## High-Pressure Photoluminescence Measurements

$NiPS_3$ crystallises in a monoclinic structure with the space group $C_{2h}^3$ and an ABC stacking order[10] [See Fig. S1]. High-quality $NiPS_3$ single crystals were grown using the chemical vapour transport method, displaying the well-characterised X-ray diffraction (XRD) pattern and a distinctive black, shiny appearance (Fig. 1a). Consistent with previous reports[6], a sharp PL peak is observed in our $NiPS_3$ crystals at low temperature (Fig. 1b), accompanied by broad-and-weak neighbouring peaks induced by phonon sidebands[7]. The primary PL peak is centred at ~1.4753 eV, closely matching the previously reported value of ~1.4756 eV[6]. The schematic of the high-pressure PL measurement setup is shown in Fig. 1c, with the actual high-pressure cell shown in Fig. S16. PL spectra of $NiPS_3$ were collected at 7 K under pressures up to 1.5 GPa, as displayed in Fig. 1d and 1e. The data reveal rapid suppression of the PL peak under pressure as low as 0.4 GPa, with the peak vanishing completely at ~1.5 GPa, as marked by the pink bar. Before the PL's extinction, its polarization remains similar under high pressure (see Fig. S11). To confirm the reversibility of this phenomenon, we reduced the applied pressure after suppression and observed that the PL signal fully recovered (Fig. S2), indicating that the suppression is reversible.

Due to the coexistence of the sharp PL peak with several neighbouring broad and weak peaks, we employ a multi-Lorentzian fitting procedure[7] to isolate the pure excitonic peak from the raw spectra. Figure S3 shows that the cumulative curve, obtained by summing all the fitted components, matches well with the experimental PL spectrum, validating our fitting approach. Notably, the extracted PL peak parameters are robust against variations in the fitting procedure and parameters.

Figure 2a presents the pressure-energy mapping of the PL intensity derived from the extracted pure PL peaks at 7 K under varying pressures. The pressure dependence of the PL amplitude, PL energy, and full-width-at-half-maximum (FWHM) are summarised in Fig. 2b-d, respectively. The error bars in these figures are negligible and smaller than the symbol size. Upon increasing pressure up to 1.2 GPa, the PL amplitude decreases monotonically by approximately 27-fold. Concurrently, the PL peak undergoes a redshift from ~1.4753 to ~1.4614 eV, while the FWHM increases gradually from ~0.48 to ~7.07 meV.

## High-Pressure Raman Spectroscopy and X-ray Absorption Spectroscopy (XAS)

A striking feature of the pressure-dependent PL data is the exciton's extreme sensitivity to a narrow pressure range. For instance, the PL peak undergoes significant changes at pressures as low as 0.4

GPa and is fully suppressed at ~1.5 GPa. Notably, this pressure range corresponds to expected lattice changes of only ~0.5%[11]. Furthermore, these small pressures are an order of magnitude lower than the critical pressures for structural (~15 GPa) and electronic (~20 GPa) transition in $NiPS_3$[11], and approximately 100 times lower than its bulk modulus of ~74.9 GPa[12]. A recent study[47] estimated the Grüneisen parameter of $NiPS_3$ to be ~1.6, suggesting that the total energy change per unit cell under 1 GPa of pressure is less than 1%. This analysis indicates that the electronic ground state of $NiPS_3$ is unlikely to be significantly altered within the current pressure range. Consequently, structural and electronic transitions[48-52] can be excluded as the primary explanations for the observed PL suppression.

We conducted high-pressure Raman spectroscopy and XAS measurements to further support our conclusion. First, the Raman spectra of $NiPS_3$ provide compelling evidence that the strong suppression of magnetic excitons is not due to structural transitions. As depicted in Fig. 3a, $NiPS_3$ exhibit several dominant Raman peaks near ~182, 259, 389, and 592 cm$^{-1}$, corresponding to the $E_g^{(2)}$, $A_{1g}^{(1)}$, $A_{1g}^{(2)}$, $A_{1g}^{(3)}$ Raman-active phonon modes[10], respectively. These phonon modes show minimal changes within 1 GPa, with only slight shifts toward higher wavenumbers at higher pressures (e.g., 3 and 7 GPa). Such observation is more evident in Fig. 3b, where the extracted Raman peak positions show little shifts within 1 GPa. This confirms that no structural transition or significant changes/contraction in lattice constants occur within the pressure of 1.5 GPa. These results are consistent with earlier findings that structural transition in $NiPS_3$ occurs only at pressures exceeding ~15 GPa[11]. Therefore, the sensitive PL modulation observed at lower pressures cannot be attributed to structural transitions at higher pressures.

Second, $NiPS_3$ is characterised by ligand-to-metal charge transfer between nickel 3*d* and sulfur *p* orbitals[5], which determines the electronic ground state of $Ni^{2+}$ ions as a predominantly $d^9\underline{L}$ multiplet configuration with contributions from the mixed $d^8$ state[6]. To investigate whether pressure-induced changes in the electronic structure could explain the exciton modulation, we performed the XAS measurement under pressure. As shown in Fig. 3c, the XAS spectra remain largely unchanged up to 3.2 GPa, retaining key features, including the pre-edge feature at 8.333 keV (▽) and three additional main σ* satellites (A, B, C). These minor changes indicate that the local electronic structure remains stable under low-pressure conditions. In contrast, significant spectral changes are observed at 29.5 GPa (Fig. 3c), likely due to a phase transition to a three-dimensional structure. The enhanced spectral dispersion at this pressure may signal an insulator-metal transition, consistent with previous reports[53].

**Temperature-Pressure (T-P) Quantum Phase Diagram**

To comprehensively understand the evolution of entangled excitons, we performed PL measurements across the temperature-pressure (T-P) parameter space. The applied pressure changes naturally in our high-pressure setup as the temperature increases during the warming process (see Fig. S3 for an example). Adjusting the setup at each temperature to maintain a fixed pressure would be time-consuming and could introduce errors. Instead, we recorded the pressure at each temperature during the warming measurement sequence without correcting for the temperature-induced pressure variations.

Figures S4-S9 display six T-P-matrix measurement sequences, including raw data and detailed spectra. Based on these measurements and subsequent analyses, we constructed the T-P quantum phase diagram of $NiPS_3$, shown in Fig. 4, using the PL amplitude as a metric. Additionally, high-pressure NMR measurements (Fig. S17) provided insights into the Neel temperature ($T_N$) of $NiPS_3$ under varying pressures. The white circles in Fig. 4 indicate the extracted $T_N$, which increases roughly linearly with pressure up to ~2 GPa (dashed gray line as a guide to the eye). The NMR results reveal that increasing pressure strengthens the magnetic ground states, which excludes a simple hypothesis that increasing pressure weakens magnetic order and thus suppresses the magnetic exciton. Such exclusion of magnon contribution can be further strengthened by the high-pressure low-frequency Raman spectra in Fig. S10, where the magnon Raman mode shows very little shift below 2.1 GPa before being heavily suppressed at around 5.2 GPa. It directly demonstrates, again, that the rapid suppression of PL below 1 GPa cannot be attributed to magnon contributions.

Finally, we identified two distinct quantum states: the bright state, characterised by the survival of sharp, coherent excitons, and the dark state, where these excitons are eliminated. The T-P evolution of the PL energy and FWHM is summarised in Fig. S12, derived from experimental data. Consistent with the trends observed in Figs. 1-2, the PL peak redshifts and the FWHM broaden with increasing pressure at a fixed temperature.

## First-Principles Many-Body Calculations

We performed first-principles many-body calculations to investigate the observed pressure sensitivity of excitonic behaviour[13,54,55]. Using the generalised Kohn-Sham approaches and the Bethe-Salpeter equation (BSE), we computed the optical spectra of $NiPS_3$. Figure S13a-c presents the calculated imaginary part of the dielectric function with the optical transition oscillator strengths (vertical solid lines) for bulk $NiPS_3$ along various directions in the crystal under pressure of 0 GPa (red solid lines), 1 GPa (green solid lines), and 2 GPa (blue solid lines). The excitons near 0.8 and 1.5 eV produce weak peaks superimposed on the dielectronic functional tail. Notably, the peaks at ~1.5 eV in the zz component are significantly weaker than their in-plane counterparts, highlighting a pronounced anisotropy in exciton excitations. With increasing pressure, these low-energy excitons exhibit a monotonic blue shift, as do their corresponding peak in $\mathrm{Im}\varepsilon_{ii}$. In contrast, the exciton states just below and above the electronic band gap remain largely unchanged (Fig.

S13d).

While the electronic band gap and exciton states show minor variations, the basic electronic ground states are nearly unaltered under small pressures below 2 GPa. These findings are consistent with our high-pressure Raman and XAS measurements and earlier discussions. Interestingly, while our first-principles calculations predict a blue shift of the primary excitonic peak below the band gap with increasing pressure, it contrasts with the experimentally observed redshift. This discrepancy suggests that the experimentally observed pressure-induced exciton modulation cannot be explained within the conventional framework of these calculations.

To explore an alternative mechanism, we examined whether a larger momentum separation between the valence band maximum (VBM) and the conduction band minimum (CBM) could account for the exciton suppression, as proposed in previous studies. Figure S14 shows the calculated momentum distribution of the VBM and CBM within the experimental range. Our results indicate that neither VBM nor CBM shifts significantly in momentum space within the experimental pressure range, and their connecting momentum remains unchanged. This finding rules out momentum separation as a viable explanation for the exciton fragility observed under small pressures. Our first-principles many-body calculations indicate no discernible variation in electronic or excitonic properties through conventional mechanisms within the current experimental pressure range. This strongly suggests the involvement of an alternative, unconventional mechanism driving the observed excitonic behaviour under small pressures.

## Discussion and Conclusion

We report a dramatic pressure dependence of magnetic excitons in $NiPS_3$, where excitonic emissions progressively weaken, redshift and are completely suppressed at a small pressure of 1.5 GPa. All experimental data and first-principles calculations consistently indicate that the ground states remain virtually unchanged within the pressure range. Moreover, conventional many-body first-principles calculations predict contrasting blueshifts in the optical channels, which sharply contradict the observed redshift. Meanwhile, our pressure experiment provides definitive, clean experimental evidence, going beyond previous chemical substitution studies [see Fig. S19 for a summary of the distinctions between pressure and doping results]. First, because PL is quenched while $T_N$ increases, the disappearance of the magnetic exciton cannot be attributed to a simple weakening of the antiferromagnetic ground state. Second, because pressure contracts the lattice while substitution expands it, the effect cannot be reduced to a simple lattice-expansion mechanism. Finally, because our pressure experiment is reversible and in-situ, the quenching is not a consequence of accumulated chemical disorder. These results establish that the optical brightness of the magnetic exciton is controlled by a higher-order spin-charge-lattice/orbital correlation that is extraordinarily fragile under clean perturbation. We propose three plausible physical scenarios to account for the novel experimental observation. These proposals are based on our experimental and theoretical observations in $NiPS_3$ under pressure, the absence of structural, magnetic, and electronic phase transitions, and on the spin-charge coupling in $NiPS_3$, where the spin, charge, and lattice degrees of freedom are intimately tied[6,43].

Figure 5a,b presents the first mechanism to explain the pressure-driven suppression of the optical transition for the magnetic exciton, i.e., the formation of the paired ZRS state (PZRS). During the magnetic exciton transition, there is always a channel for the ZRT to ZRS transition[6,56]. Importantly, it is known that there is a local spin-charge coupling for the S(*p*) electron[6]. The S(*p*) electrons shared by Ni atoms having antiparallel (parallel) spin are spin-unpolarized (polarized). The spin-unpolarized (polarized) S sites have more (less) electrons due to the Hund's coupling of the S(*p*) electron, as shown by solid (dashed) ellipses in Fig. 5a,b. From this spin-charge coupling mechanism, if a PZRS state forms in the nearest Ni sites in the FM chain direction, the number of the spin-polarized and the spin-unpolarized S($p$) atoms are conserved, satisfying the charge neutrality for the stable state.

On the other hand, the single ZRS (SZRS) state does not obey the charge neutrality from the spin-charge coupling. This charge-neutrality condition could drive a stable PZRS state between the two separated SZRS states, as shown in Fig.5a and 5b. As illustrated in Fig. 5a, two Ni sites in the PZRS state share the S atoms, thereby inheriting quantum entanglement between the two ZRS states, and acquiring a spin-character-conserving component in the PZRS state during excitation from the ZRT ground state. This spin-character-conserving component in the PZRS state implies that the PZRS state is optically bright, as depicted in Fig. 5a. Our first scenario is that pressure could be a critical perturbation for the paired state, making the optically bright PZRS state dissociate into the two optically dark SZRS states under pressure.

Fig. 5c illustrates the second mechanism to explain the pressure-driven suppression of the optical transition for the magnetic exciton: crystal-field-dependent spin-orbit coupling effects on the Ni($d^8$) multiplets. For the $d$-$d$ transition channel of the magnetic exciton, the spin value is not conserved for the optical transition. However, finite spin-orbit coupling could relax the spin conservation rules[56]. The half-filled $e_g$ manifold in the Ni($d^8$) state has a quenched angular momentum, separated from the $t_{2g}$ manifold by the octahedral crystal field of 10Dq around ~1 eV[6]. The Ni($d$) spin-orbit coupling constant is around 80 meV[6]. The relative energy of 10Dq and the spin-orbit coupling constant imply that spin-orbit coupling effects in the Ni($d^8$) multiplet are in the perturbation regime for the finite optical transition matrix element of the magnetic exciton. Thus, our second scenario indicates that the spin-orbit coupling, which relaxes the spin conservation rule, is fragile under pressure. The pressure-driven reduction of the Ni-S distances enhances the 10Dq and could suppress the optical transition matrix, as shown in Fig. 5c.

Fig. 5d depicts the third possible mechanism to explain the pressure-driven suppression of the optical transition for the magnetic exciton: the pressure-driven hardening of the inversion-symmetry-breaking component of the $A_{1g}^{(1)}$ phonon associated with the zigzag AFM configurations. It is known that the magnetic exciton in the optical transition is dressed by the $A_{1g}^{(1)}$ phonon[8,43]. Interestingly, the $A_{1g}^{(1)}$ phonon, in the presence of the AFM state, breaks the inversion symmetry at the Ni sites[43]. This inversion symmetry breaking is demonstrated in Fig. 5d. The S atoms connecting the FM Ni spins have a stronger covalent bonding (smaller displacement from $A_{1g}^{(1)}$) than that in the S atoms connecting the AFM Ni spins (larger displacement from $A_{1g}^{(1)}$), respectively. This spin-order-driven breaking of inversion symmetry for the $A_{1g}^{(1)}$ phonon mode inherits the spin-character-conserving component in the ZRS state. This component is dressed by the $A_{1g}^{(1)}$ phonon in the excitation from the ZRT ground state. We propose the third scenario: under pressure, the reduction of the Ni-S distance hardens the inversion-symmetry-breaking component of the $A_{1g}^{(1)}$ phonon. This suppression of the inversion-symmetry-breaking associated with the AFM configuration could reduce the spin-character-conserving component of the ZRS state dressed by the $A_{1g}^{(1)}$ phonon, explaining the fragile nature of the magnetic exciton under pressure.

We present these scenarios in order of physical plausibility based on our current experimental constraints. They are not completely mutually exclusive, because exciton pairing, crystal-field-controlled spin-orbit mixing, and symmetry-breaking may act cooperatively in determining the optical brightness of the magnetic exciton. Distinguishing their relative contributions will require future microscopic probes, but the present pressure results already impose stringent constraints: any viable theory must explain why the exciton is quenched while the lattice is slightly compressed, the electronic structure remains stable, and the antiferromagnetic transition temperature increases.

We stress several novel features of $NiPS_3$: the very small charge-transfer energy, the charge-spin coupling, the spin-charge-orbit correlation, and finally, the effective one-dimensional charge stripe associated with the zigzag AFM order. Finally, we elucidate the new temperature-pressure quantum phase diagram with two distinguished quantum states. Our findings unravel the abundant exotic properties of entangled excitons in $NiPS_3$ from multiple perspectives and stimulate future understanding and exploration of coherent many-body interactions in magnetic excitons.

## Methods

### Growth and structure characterization of $NiPS_3$ single crystals

High-quality $NiPS_3$ single crystals were grown by the chemical vapor transport method, as described in our previous work[6]. Ni, P, S, and $I_2$ were mixed inside an argon-filled glove box and sealed in a quartz tube under vacuum. The quartz tube was placed in a two-zone furnace with a temperature gradient of 750 ˚C (source) to 700 ˚C (sink) for 7 days. Crystal structures were characterized by single-crystal XRD.

### High-pressure technique, PL and Raman spectroscopy measurements

The pressure was generated by the symmetric diamond anvil cell (DAC) (Fig. S16), with Rhenium as the gasket material and Daphne 7373 oil as the pressure-transmitting medium. The $NiPS_3$ crystal was loaded into a symmetric DAC with a culet diameter of ~400 μm. A gasket with a ~200 μm hole was used for the sample chamber. Small ruby spheres with a diameter of ~5-10 μm were also loaded for pressure estimation based on their fluorescence signal (Fig. S18). The in-situ high-pressure photoluminescence and Raman spectroscopy measurements were performed with the Renishaw inVia Raman Microscope, equipped with a cryostat and a temperature controller of the Helium flow exchange type. A 632.8 nm laser produces the excitation.

### High-pressure XAS experiments

A symmetric DAC was prepared for the in-situ X-ray absorption spectroscopy at high pressure up to 29.5 GPa, using the same high-pressure technique above for the PL measurements. Ni K-edge x-ray absorption near-edge spectra for $NiPS_3$ were collected at the 3A beamline of the Pohang Accelerator Laboratory (PAL), Korea, using a Si(111) double-crystal monochromator with varying incident photon energy. The intensities of X-ray fluorescence signals were measured using the pin diode with a step size of 0.2 eV. The six XAS spectra were averaged to yield the final spectrum presented here. The beam size (full-width-at-half-maximum) on the target sample is 90 μm (v) $\times$ 300 μm (h).

### First-principles many-body calculations

First-principles calculations were carried out by using the pseudopotential projector-augmented wave (PAW) method[57] implemented in the Vienna ab initio simulation package (VASP)[58,59] with an energy cutoff of 400 eV for the plane-wave basis set. The GW PAW potentials released with VASP.5.4 were used. Exchange-correlation effects were treated using the SCAN meta-GGA scheme[60] in combination with the rVV10 van der Waals (vdW) density functional[61]. A $7 \times 3 \times 7$ Γ-centered k-point mesh was used to sample the Brillouin zone. The spin-orbit coupling for the Ni atom was not considered.

The crystal volume, shape, and atomic positions were relaxed such that the norm of all the forces on each atom is smaller than 0.01 eV/Å. The relaxation was initialized using the experimentally determined atomic positions and lattice parameters for the bulk C2/m (space group 12) structure[62]. A total energy tolerance of $10^{-8}$ eV was used to determine the self-consistent charge density and total energy. The response functions and exciton eigenvalue calculations followed the procedure of previous works[13,54], where we adopted the single-shot $W_0$ variant of the fully self-consistent screened interaction commonly employed in the GW approximation due to its reasonable computational performance while maintaining robust results[63]. Specifically, we calculated screened interaction at the $W_0$ level using 96 frequency points and 762 virtual orbitals with an energy cutoff of 220 eV.

To obtain excitons, the Bethe-Salpeter equation (BSE) was solved beyond the Tamm-Dancoff approximation, where coupling between positive and negative frequencies is included in the BSE

Hamiltonian. Since SCAN is constructed within the generalized Kohn-Sham scheme[64] and our recent SCAN-based study of $NiPS_3$ obtains a band gap in good accord with experimental observations[55], we avoid the GW quasiparticle corrections and directly use the generalized Kohn-Sham band energies as the eigenvalues of the electrons and holes in the BSE Hamiltonian. Furthermore, 50 conduction and 30 valence bands were included in constructing the BSE Hamiltonian.

## Data availability

Data supporting this study are available from the corresponding authors upon request.

**Acknowledgments:** We thank Naoto Nagaosa, Yingming Xie, Xiaoxiao Zhang, Jaehoon Kim, Zengming Zhang, Hongze Zhao, Feng Jin, Qingming Zhang, and Kyungwon An for their helpful discussions and experimental support. We acknowledge the CAC of KIAS for computational resources and the 3A beamline of the PAL, Korea. Work at SNU was supported by the Leading Researcher Program of the National Research Foundation of Korea (Grant No. 2020R1A3B2079375) and the National Research Foundation (NRF) of Korea (Grant No. RS-2026-25484445), funded by the Ministry of Science and ICT. MZ was supported by the Natural Science Foundation of Anhui Province (2308085QA18), and ZY was supported by the National Key Research and Development Program of China (2022YFA1602603 and 2023YFA1406102). This work was also supported by the Basic Research Program of the Chinese Academy of Sciences Based on Major Scientific Infrastructures (No. JZHKYPT-2021-08). SKL was supported by the National Research Foundation of Korea (NRF) grant (2020R1A3B2079815, Leader program). MK and YWS at KIAS were supported by KIAS Individual Grants (CG083502 and CG068702). BHK was supported by the Institute for Basic Science in the Republic of Korea through Project (No. IBS-R024-D1). HC was supported by the National Research Foundation of Korea (NRF) grants (RS-2024-00450714 and RS-2024-00441954).

**Author contributions:** KXZ, YWS, and JGP initiated and supervised the project. The samples were synthesized and characterized by KXZ, JK, HY, and YL. PL experiments were carried out by MZ and ZY. XAS measurements were performed by YHK, PP, CB, BGC, HDK, TYK, and SKL. NMR measurements were done by JH and SHB. First-principle calculations were conducted by CL and JXZ. The theoretical scenarios were made by MK, BHK, and YWS. MD and KHK participated in helpful discussions. HC provided constructive suggestions to improve the manuscript. KXZ, JGP, and YWS analyzed most of the data. KXZ, MK, YWS, and JGP wrote the manuscript with contributions from others.

**Competing interests:** The authors declare no competing interests.

**Additional information:** Supplementary information is available in the online version of this paper. Reprints and permissions information is available online at www.nature.com/reprints. Correspondence and requests for materials should be addressed to the corresponding authors.

**Supplementary Materials include:**

Figs. S1 to S19

References

## Figures

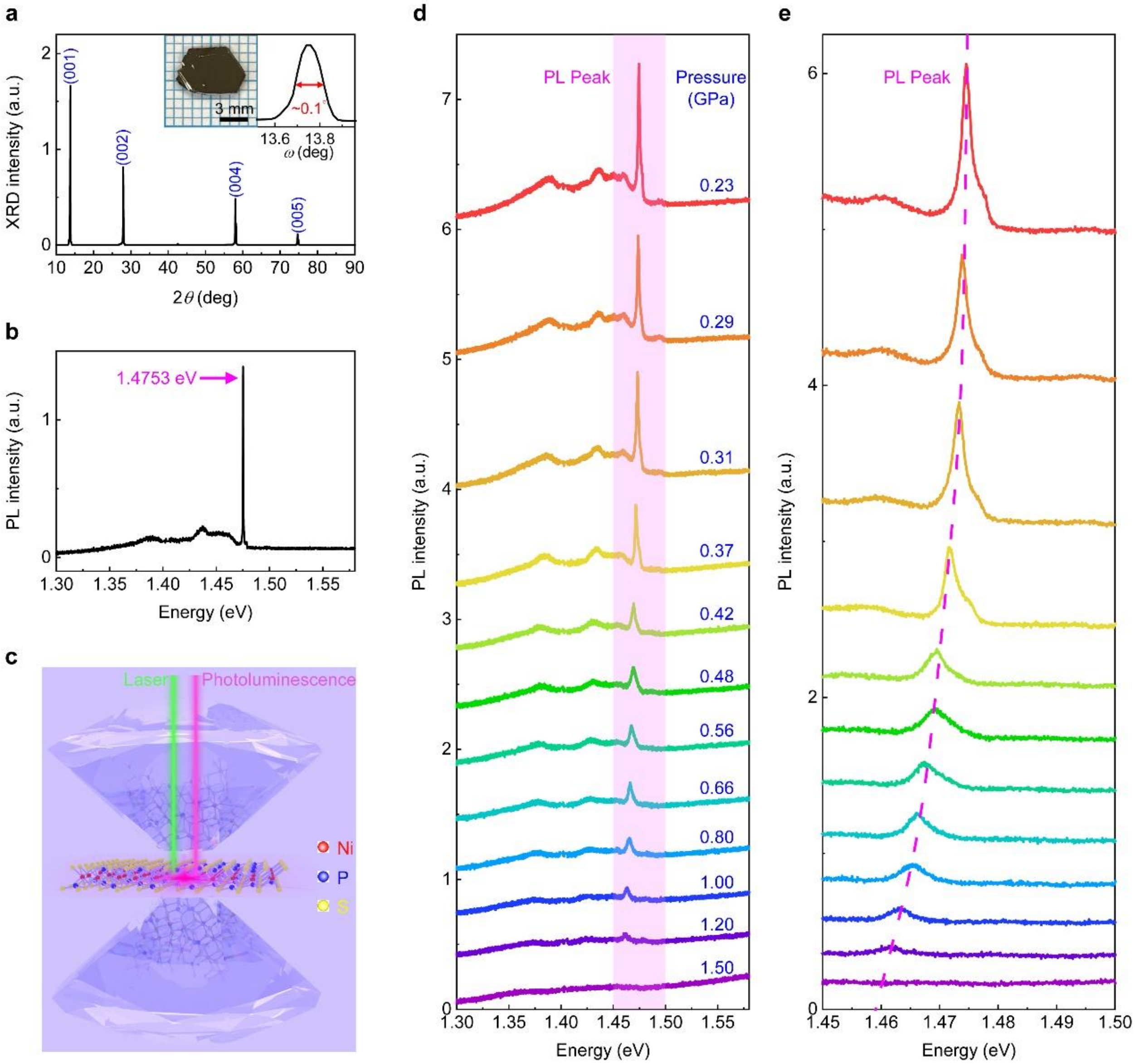

**Fig. 1 | Photoluminescence under varying pressures. a**, Optical images and XRD pattern of the $NiPS_3$ single crystal. The FWHM of the rocking curve is about 0.1°. The scale bar is 3 mm. **b**, Sharp PL peak of $NiPS_3$ at 7 K. **c**, Schematic of high-pressure PL measurement. **d**, Photoluminescence response under changing pressures up to 1.5 GPa at 7 K. The sharp PL peaks occur in the energy range of 1.45-1.5 eV, as indicated by the pink bars. These peaks are accompanied by neighboring broad-and-weak peaks induced by phonon sidebands. **e**, Blown-up picture of the spectra of the PL peaks. The PL peak is gradually suppressed with increasing

pressure, showing a decreasing PL intensity, and vanishes at ~1.5 GPa. Meanwhile, the PL peak redshifts gradually while applying increasing pressure, as depicted by the dashed pink curve.

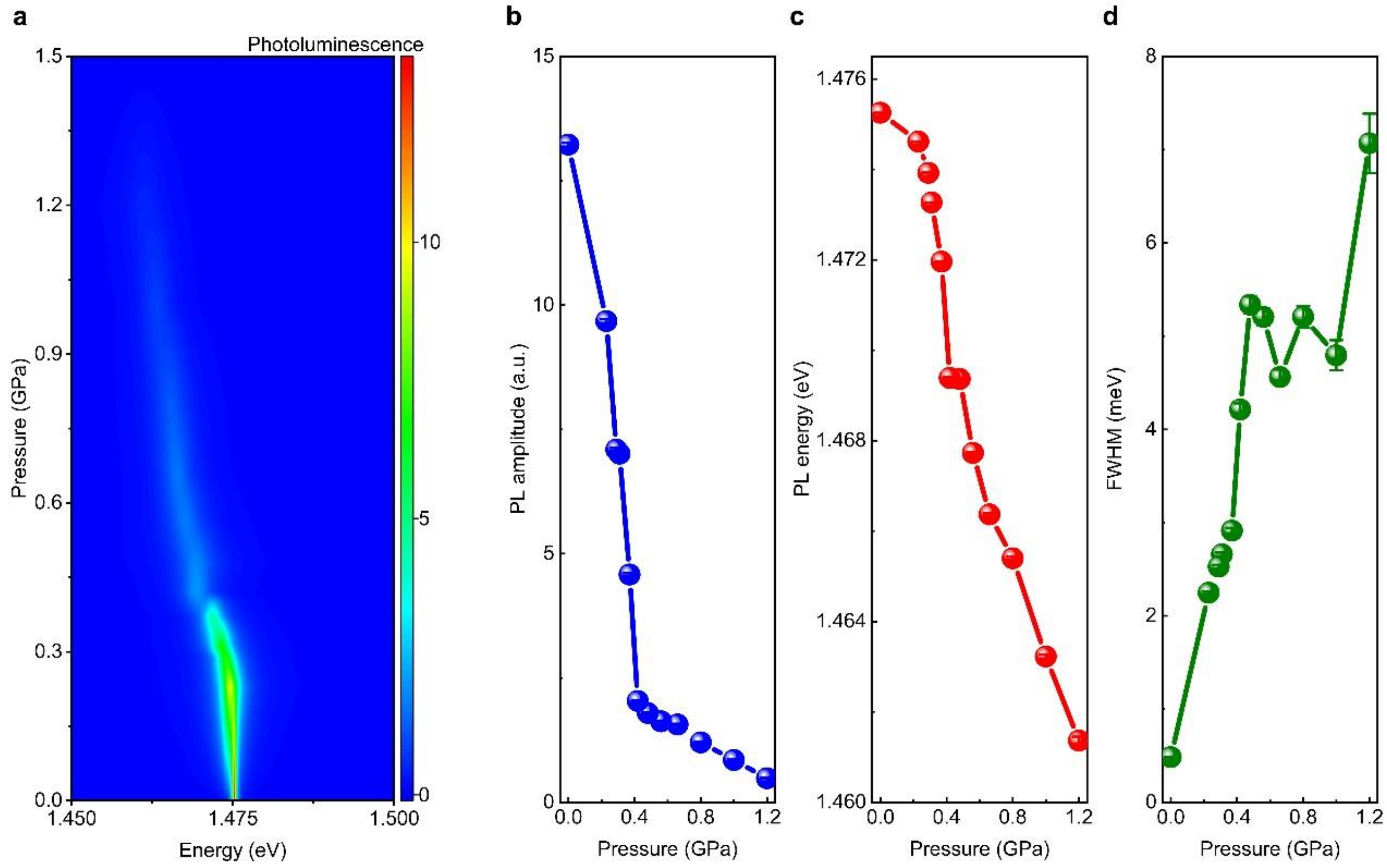

**Fig. 2 | Evolution of pure excitonic peak and its physical parameters under pressure. a**, Pressure-energy mapping of PL intensity from the extracted pure excitonic peak by the multi-Lorentzian fitting (see Fig. S3 as an example) under various pressures at 7 K. **b-d**, Evolution of PL amplitude (**b**), PL energy (**c**), and FWHM (**d**) as a function of applied pressure, respectively. The PL amplitude and energy decrease monotonically while the FWHM increases as pressure rises from 0 to 1.2 GPa.

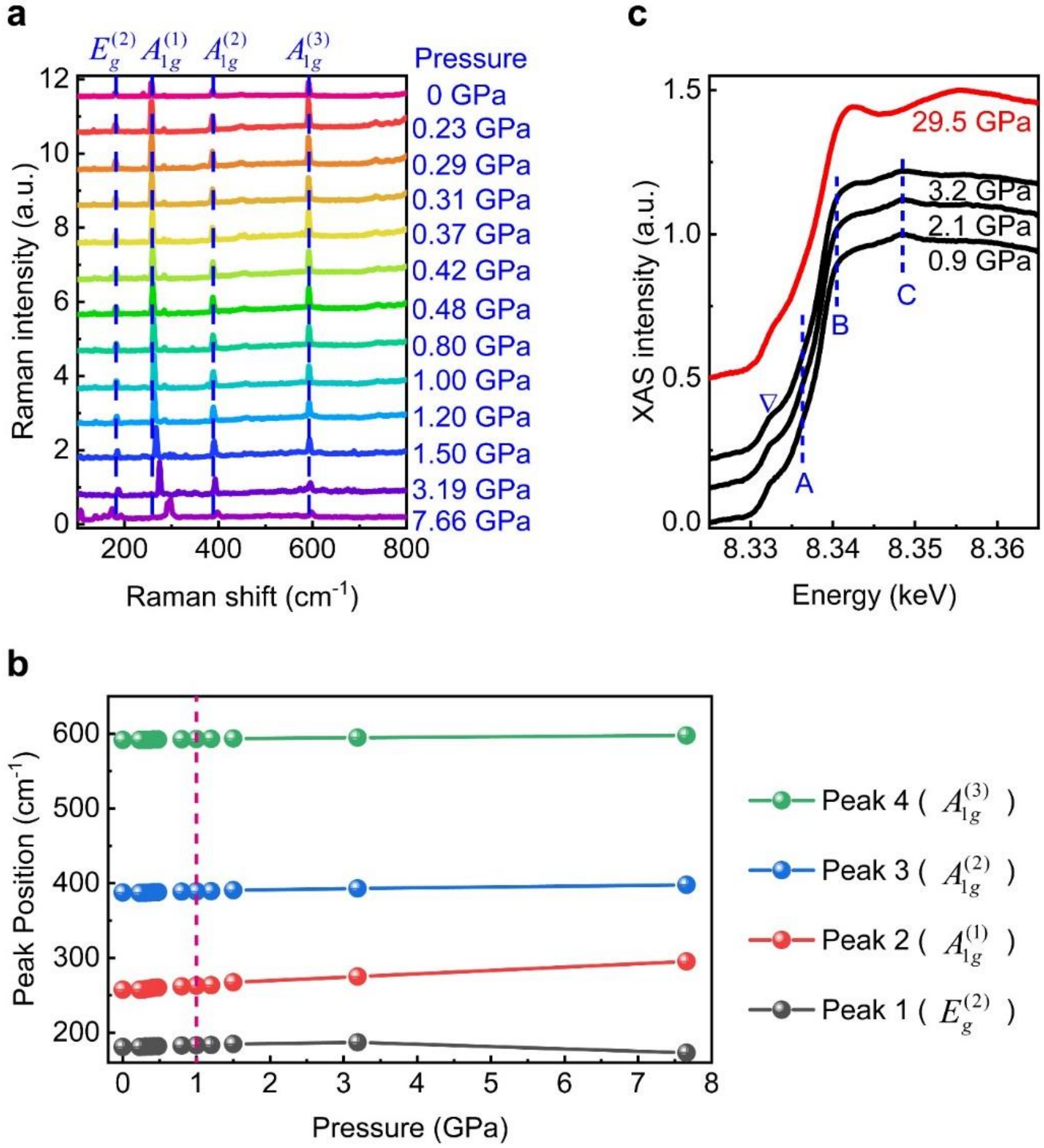

**Fig. 3 | Raman and XAS spectra under changing pressures. a**, Raman spectra of $NiPS_3$ at 7 K under different pressures. The dashed blue lines indicate the main Raman peaks near ~182, 259, 389, and 592 cm$^{-1}$ under increasing pressure, which remain little changed up to 1 GPa. The tiny change is amplified for eye guidance in (**b**). **b**, Raman peak positions as a function of pressure. They show little change up to 1 GPa, where the dramatic PL suppression occurs, ruling out structural transitions or sharp structural changes as possible explanations. **c**, XAS spectra under different pressures: 0.9, 2.1, 3.2, and 29.5 GPa, respectively. The XAS spectra within 3.2 GPa exhibit the same pattern: a pre-edge feature at 8.333 keV (▽) and three additional main σ* satellites (A, B, C), which is in stark contrast to the XAS pattern under 29.5 GPa.

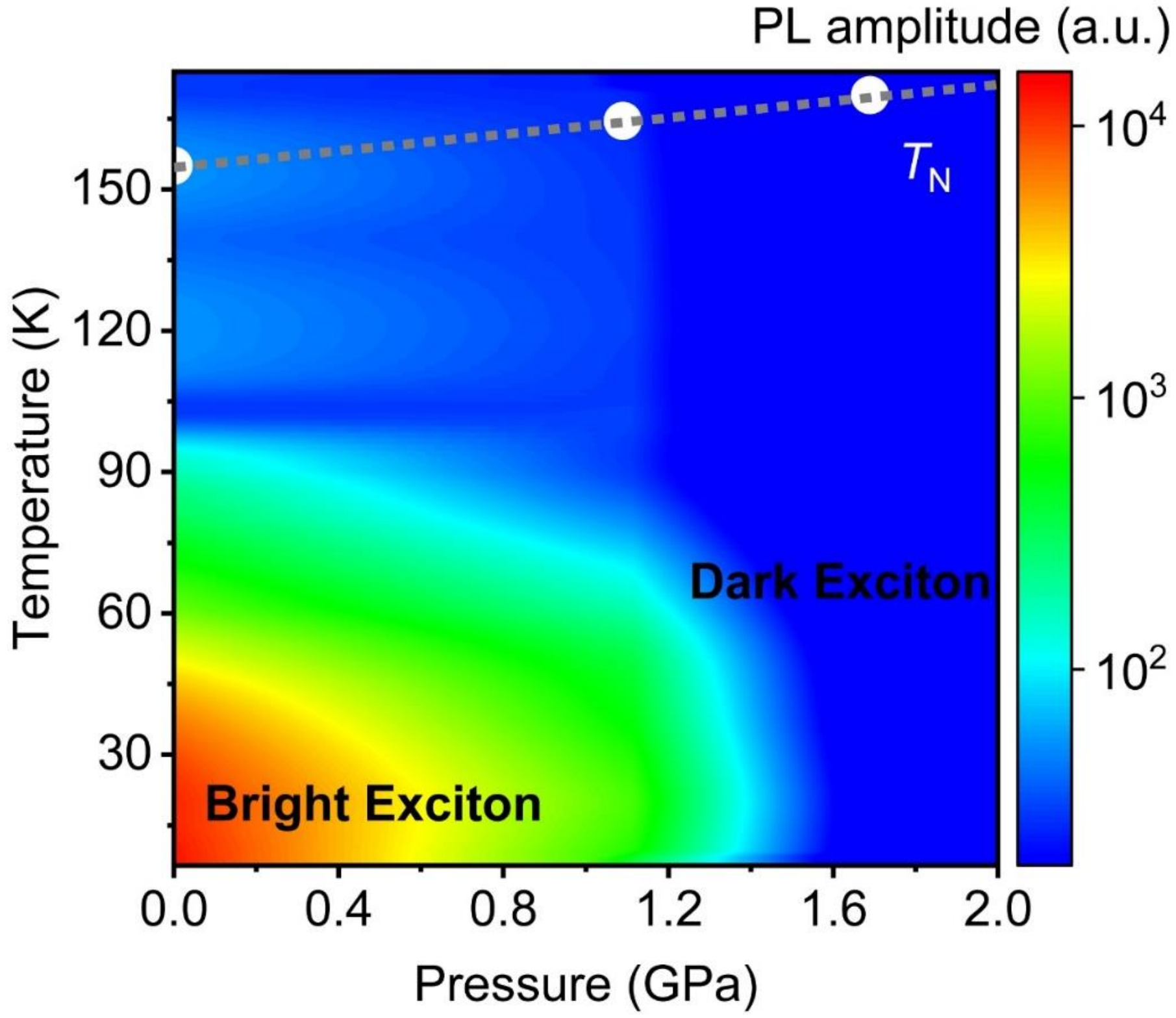

**Fig. 4 | Temperature-pressure quantum phase diagram.** An experimental T-P phase diagram of $NiPS_3$ was obtained by analyzing all the excitonic peaks in Figs. S4-S9. The dashed gray line shows a slight increase in $T_N$ (white circles) with increasing pressure, as confirmed by high-pressure NMR measurements in Fig. S17. A phase boundary can be observed between the bright and dark exciton quantum states, depicting the stable T-P region for the sharp coherent magnetic excitons. The bright exciton corresponds to a significant, sharp photoluminescence in its pristine, delicate quantum states. In contrast, the dark exciton indicates the quenching of photoluminescence by the pressure-induced magnetic exciton destruction.

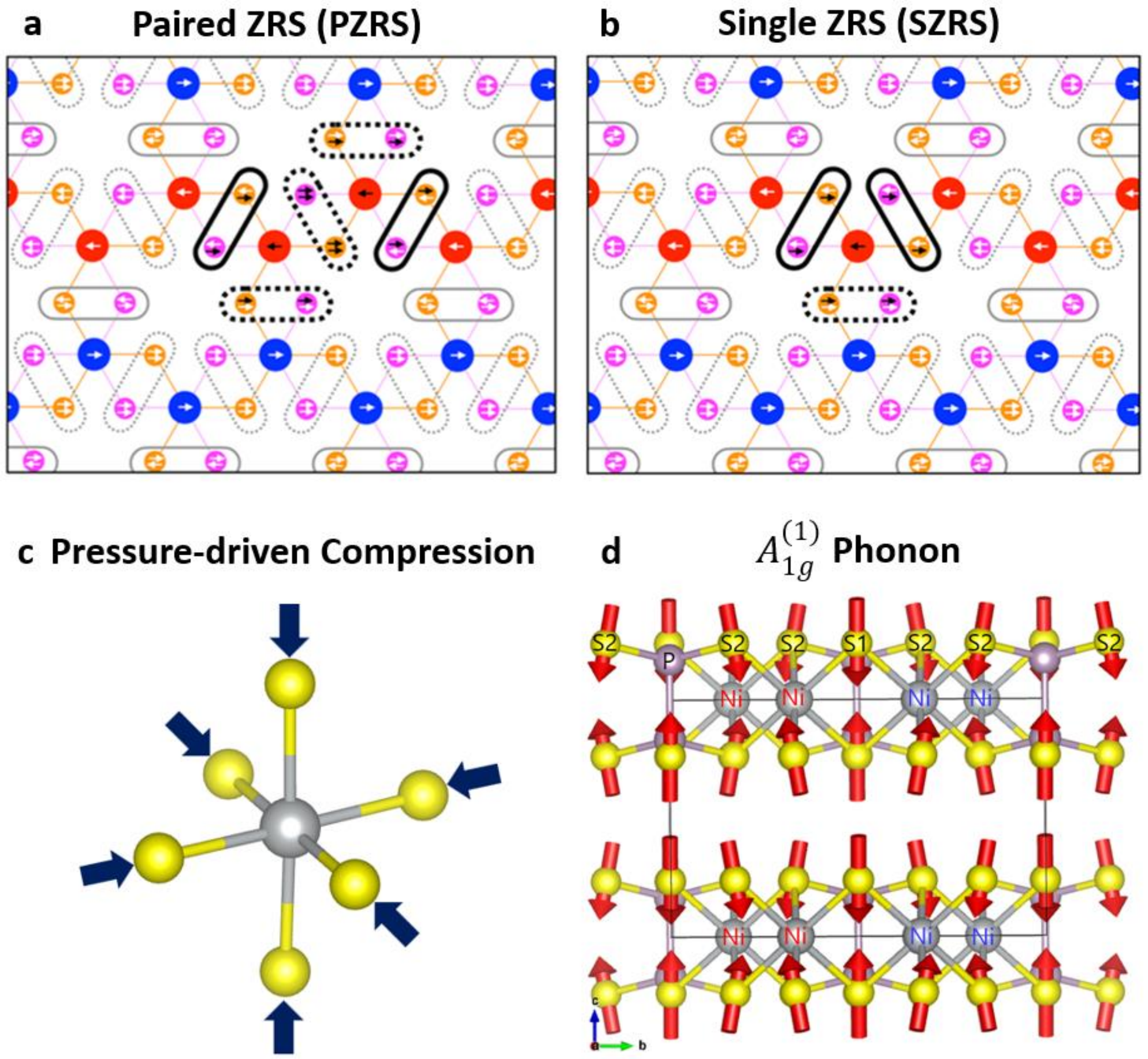

**Fig. 5 | Possible scenarios for the fragile nature of the magnetic exciton. a-b,** Concept of the paired Zhang-Rice singlet (PZRS) exciton. (a) and (b) Spin and S(*p*) charge configurations in the PZRS state and the single ZRS (SZRS) state, respectively. Top views are shown for the spin configurations of the Zhang-Rice triplet (ZRT) ground state (white arrows) and the ZRS exciton (black arrows), respectively. Dashed (solid) ellipses present spin-polarized (unpolarized) S(*p*) electrons, respectively. **c,** Pressure-driven reduction of the Ni-S distance, enhancing the crystal-field splitting of 10Dq. **d,** Inversion symmetry breaking at Ni sites from the $A_{1g}^{(1)}$ phonon excitation associated with the zigzag AFM order. Red arrows present the S atom displacement for the $A_{1g}^{(1)}$ phonon, whose magnitude is larger for the AFM connection than the FM connection.